\documentclass[preprint,review,1p,12pt]{elsarticle}
\usepackage{amssymb}
\usepackage{bm}
\usepackage{geometry}
\journal{Physics Letters A}

\begin{document}

\begin{frontmatter}

\title{Effects of Rashba spin-orbit coupling and a magnetic field on a polygonal quantum ring}


\author[label1]{Han-Zhao Tang}
\author[label1]{Li-Xue Zhai}
\author[label1,label2]{Man Shen\corref{cor1}}
\ead{shenman@semi.ac.cn}
\address[label1]{Physics Department and Hebei Advanced Thin Film Laboratory, Hebei Normal University, Shijiazhuang 050024, Hebei, People's Republic of China}
\address[label2]{SKLSM, Institute of Semiconductors, Chinese Academy of Sciences, P.O. Box 912, Beijing 100083, People's Republic of China}
\author[label1,label3]{Jian-Jun Liu\corref{cor1}}
\ead{liujj@mail.hebtu.edu.cn} \cortext[cor1]{Corresponding author}
\fntext[fn1]{No. 288, Zhufeng Street, Shijiazhuang, Hebei, People's Republic of China, 050035.} \fntext[fn1]{Tel:
+86-311-66617006; Fax: +86-311-66617006}
\address[label3]{Physics Department, Shijiazhuang University, Shijiazhuang, Hebei 050035, People's Republic of China}

\begin{abstract}
Using standard quantum network method, we analytically investigate the effect of Rashba spin-orbit coupling (RSOC) and a magnetic field on the spin transport properties of a polygonal quantum ring. Using Landauer-B\"{u}ttiker formula, we have found that the polarization direction and phase of transmitted electrons can be controlled by both the magnetic field and RSOC. A device to generate a spin-polarized conductance in a polygon with an arbitrary number of sides is discussed. This device would permit precise control of spin and selectively provide spin filtering for either spin up or spin down simply by interchanging the source and drain.
\end{abstract}


\end{frontmatter}


\section{Introduction}
In recent years, there has been great interest in quantum computation and in applications of low dimensional semiconductor devices in which spin-dependent transport phenomena are apparent \cite{1,2}. Datta and Das proposed a spin-field-effect transistor device for controlling the spin of an electron \cite{3}. When an electron is transmitted from the source to the drain in this device, the spin precession and the spin-dependent phase of the electron could be controlled by the Rashba spin-orbit coupling (RSOC), which could be controlled by a perpendicular electric field \cite{3}. In addition, both the Aharonov-Bohm (AB) \cite{4} and Aharonov-Casher (AC) \cite{5} effects have been studied for quantum rings both experimentally and theoretically and may also be useful for controlling the spins of electrons \cite{6,7,8}.

Recently, high quality semiconductor rings have been made, and have attracted significant attention because of the interesting interference phenomena that arise due to their geometric structure \cite{9}. The spin-dependent conductance of a circular quantum ring system with RSOC have been studied by Aeberhard $et~al$. \cite{10}. Besides the work on circular ring models \cite{9,10,11,12}, other geometries have also been proposed and studied \cite{13,14,15,16,17,18,19,20,21}. Very recently, Bercioux $et~al$. \cite{15} have studied the localization of the electron wave function in a quantum network with both RSOC and an external magnetic field. The spin transport properties of a regular polygon with RSOC have also been studied \cite{18}. It is now known that an electron in a ring with RSOC may have a spin-dependent phase that can be controlled by an adjustable gate voltage. Likewise, the transmitted electrons in a polygon should also show AB oscillation phenomena which can be controlled by a magnetic flux. Based on our previous work \cite{18}, we have constructed a theoretical model to investigate spin transport properties in polygonal quantum rings with both RSOC and a magnetic field, concentrating on AB oscillations and the breaking of time reversal symmetry by the magnetic field. Subsequently, we use this model to propose an effective spin filter based on these concepts.

In this paper, using a standard quantum network method, we analytically study the spin-dependent conductance of electrons through a polygonal ring. This paper is composed as follows. We present the theoretical framework and derive appropriate formulas in Sec. \uppercase \expandafter {2}. Sec. \uppercase \expandafter {3} presents our results and analyses. Sec. \uppercase \expandafter {4} is devoted to a summary of our work.

\section{Model and formulas}

\begin{figure}
\centering
\includegraphics[scale=0.55]{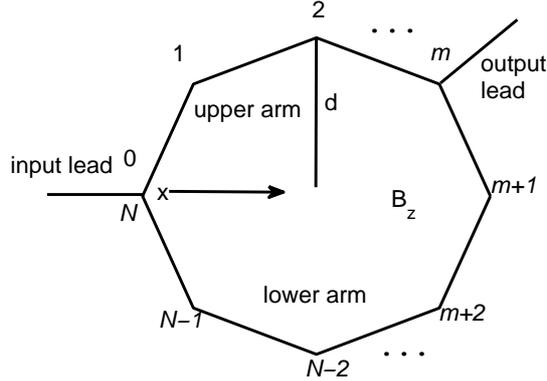}
\caption{\label{fig_1} Schematic diagram of the one dimensional (1D) polygonal quantum ring coupled to two leads. The RSOC and a perpendicular magnetic field $B_{z}$ are applied to both the upper arm (from node $0$ to node $m$) and to the lower arm (from node $m+1$ to node $N$).}
\end{figure}
A regular polygon with $N$ segments coupled to input and output lead is shown in Fig. 1. The magnetic field and RSOC are not considered in the leads. The output lead can be located at either a symmetric or asymmetric vertex with respect to the $x$ axis. For simplification of the analysis, we use the same approximation as in Ref. 18 and Ref. 22 for the segment of the quantum ring.

\
The Hamiltonian on a segment lying in an arbitrary direction $\bm{\gamma}$ and the wave functions on the input/output lead can be written in the following form \cite{23,24}:
\begin{equation}
  \hat{H}=\frac{(p_{\gamma}+qA_{\gamma})^{2}}{2m^\ast}+\frac{\hbar{k_{so}}}{m^\ast}(p_{\gamma}+qA_{\gamma})(\bm{\sigma}\times \textbf{z})\cdot\bm{\gamma},\label{Eq1}
\end{equation}
\begin{equation}
  \psi_{in}=Ae^{ik_{0}r}+Be^{-ik_{0}r},\psi_{out}=Ce^{ik_{0}r},\label{Eq2}
\end{equation}
where $A_{\gamma}$ is the vector potential and the meaning of other physical quantities are as in to Ref. 18. We neglect the Zeeman splitting introduced by the magnetic field \cite{15}. The wave function of one segment can be expressed as:
\begin{eqnarray}
\Psi_{\alpha\beta}(l)&=&\frac{e^{-if_{\alpha\l}}e^{ik_{so}l(\bm{\sigma}\times\bm{z})\cdot\bm{\gamma}_{\alpha\beta}}}{sin(kl_{\alpha\beta})}\{sin[k(l_{\alpha\beta}-l)]\Psi_{\alpha}\nonumber\\
& &  +sin(kl)e^{if_{\alpha\beta}}e^{-ik_{so}l_{\alpha\beta}(\bm{\sigma}\times\bm{z})\cdot\bm{\gamma}_{\alpha\beta}}\Psi_{\beta}\}.\label{Eq3}
\end{eqnarray}
Because the RSOC appears in the form of the exponentials that contain Pauli matrices in Eq. (3), there are existing methods that allow us to study quantum network problems with RSOC \cite{25,26}. The magnetic field gives rise to the phase factors
\begin{eqnarray}
\textrm{exp}\{-if_{\alpha\beta}\}&=&\textrm{exp}\{-i\frac{2\pi}{\phi_{0}}\int_{\alpha}^{\beta}\bm{A}\cdot d\bm{l}\}\nonumber\\
& = &\textrm{exp}\{\frac{i\pi dB_{z}l}{\phi_{0}} \sin(\frac{N-2}{2N}\pi)\},\label{Eq4}
\end{eqnarray}
where $B_{z}$ is the magnetic field intensity, and $\phi_{0}=h/e$ is the flux quantum. We define $\phi=\int_{\alpha}^{\beta}\bm{A}\cdot d\bm{l}=-\frac{1}{2}dB_{z}l\sin(\frac{N-2}{2N}\pi)$. The phase due to the magnetic field is described by the exponentials that contain the AB magnetic field intensity in Eq. (4), which is the key step in generalizing existing methods for studying quantum networks in the presence of a magnetic field. With the help of the Griffith boundary conditions, we can obtain the wave function of the whole quantum network \cite{9,27}. To solve for the spin-dependent conductance, we separately calculate expressions for $\psi_{n}^{(+)}$ and $\psi_{n}^{(-)}$, the wavefunctions for electrons in the upper arm and lower arm, respectively.
\begin{eqnarray}
\prod_{i=0}^{n}e^{a_{i}k_{so}l}\psi_{n}^{(+)}=c_{1}e^{in(f_{\alpha\beta}+kl)}+c_{2}e^{in(f_{\alpha\beta}-kl)},\label{Eq5}
\end{eqnarray}
\begin{eqnarray}
\prod_{i=0}^{n}e^{b_{i}k_{so}l}\psi_{n}^{(-)}=d_{1}e^{in(f_{\alpha\beta}+kl)}+d_{2}e^{in(f_{\alpha\beta}-kl)}.\label{Eq6}
\end{eqnarray}
Thus, we can obtain the transmission probability amplitude:
\begin{equation}
C=\frac{2i\frac{k_{0}}{k}\sin(kl)a}{a^{-1}a-[b-e^{-2if_{\alpha\beta}}\frac{\sin[(N-m-1)kl]}{\sin[(N-m)kl]}][b-e^{2if_{\alpha\beta}}\frac{\sin[(N-m-1)kl]}{\sin[(N-m)kl]}]}A        ,\label{Eq7}
\end{equation}
where\begin{eqnarray*}
a=e^{inf_{\alpha\beta}}\frac{\sin(kl)}{\sin(mkl)}e^{-(a_{m}+\cdots+a_{1})k_{so}l}+e^{i(N-n-2)f_{\alpha\beta}}\frac{\sin(kl)}{\sin[(N-m)kl]}e^{-(b_{N-m}+\cdots+b_{1})k_{so}l}, \end{eqnarray*}
\begin{eqnarray*}
b=2\cos(kl)-\frac{\sin[(m-1)kl]}{\sin(mkl)}-i\sin(kl)\frac{k_{0}}{k},
\end{eqnarray*}
\begin{eqnarray*}
a_{n}=\left(\begin{array}{ccc}0 & -e^{-i\theta_{n}^{+}} \\ e^{i\theta_{n}^{+}} & 0 \\\end{array}\right),~\theta_{n}^{+}=\frac{(N-2)\pi}{2N}-\frac{(n-1)2\pi}{N}~\textrm{and}~ b_{n}=a_{n}^{\ast}.
\end{eqnarray*}
with the help of the Landauer-B\"{u}ttiker conductance formula \cite{27,28}, the conductance in the polygonal quantum ring can be obtained as:
\begin{equation}
G=\frac{e^{2}}{h}\sum_{\sigma=\uparrow,\downarrow}|T_{\sigma}|^{2}=G_{\uparrow}+G_{\downarrow}.\label{Eq8}
\end{equation}

\section {Results and discussions}

\begin{figure}
\centering
\includegraphics[scale=0.5]{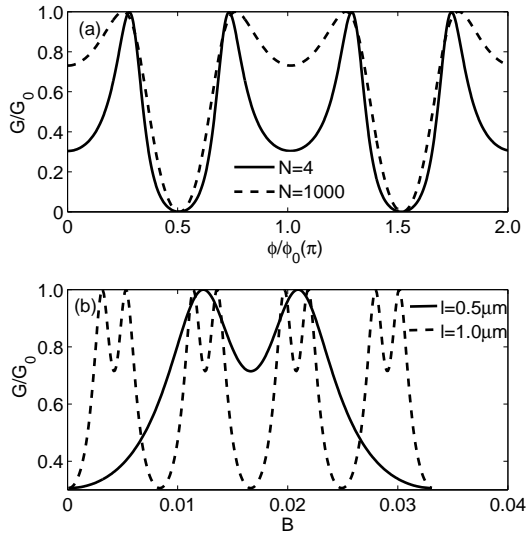}
\caption{\label{fig_2} Total conductance (a) as a function of magnetic flux for different values of $N$ in the symmetric case and (b) as a function of the magnetic field intensity for different segment lengths $l$ for $N=4$, $m=3$. We use $G_{0}=2e^{2}/h$ as the unit of conductance.}
\end{figure}
In this section, we concentrate on the physics of Eq. (7) by numerically analyzing the spin-dependent conductance. There are two ways to control the phase of electrons transmitted in the polygon. One is through the AB magnetic flux which is described by Eq. (4), and the other is through the RSOC which can be regulated by the gate voltage. In Fig. 2(a), we plot the conductance of a polygon for different numbers of segments ($N$) to describe the effect of the magnetic field when the incident electrons are unpolarized. The magnetic phase factor in the exponentials of Eq. (4) influences the phase of electrons in the polygon. The conductance is periodic in the magnetic flux for $\phi_{0}$ which is typical for the AB effect in a polygon. For the symmetric case ($m=1/2 N$), we can note that the conductance depends on the number of polygon segments, which is caused by the difference of the magnetic flux $\phi$ for different polygon areas. As the number of segments increases, the polygon approaches a circle, and $\phi=-\frac{1}{2}dB_{z}l\sin(\frac{N-2}{2N}\pi)$ approaches $\phi=-\frac{1}{2}dB_{z}l$. Therefore the geometrical factor $\sin(\frac{N-2}{2N}\pi)$ determines the phase difference between different polygonal rings and a circular ring.
\

For an asymmetric diamond ring, ($N=4$), the conductances for different segment lengths are shown in Fig. 2(b). The segment length for $l=0.5~\mu m$ is half that for the $l=1~\mu m$ case, and the distance from the center of the polygon to vertex ($d$) should also be half that of the $l=1~\mu m$ case. Since the period of the AB oscillation is $\phi_{0}$. Thus, the period of the magnetic field intensity for the $l=1.0~\mu m$ case should be quadruple that of the $l=0.5~\mu m$ case. Therefore, the conductances depend on the shape of polygonal quantum ring in the presence of a magnetic field, which is different from the conductance of a polygon with only RSOC.\cite{18} In addition, because of the AB phase interference between electrons traveling in the clockwise and counterclockwise directions, the conductance $G_{total}$ shows oscillations. When $k_{so}=0$, spin polarization is absent and the conductance can be regulated by the AB magnetic flux.

\begin{figure}
\centering
\includegraphics[scale=0.5]{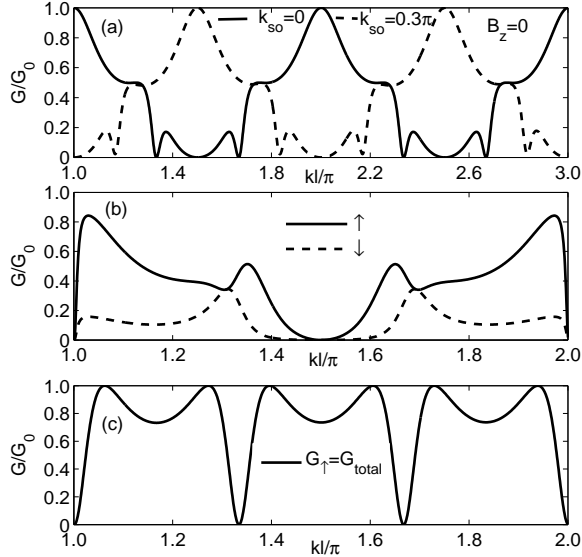}
\caption{\label{fig_3} (a) Total conductance as a function of wave-vector in a hexagonal ring. Spin-dependent conductance $G$ versus the wave-vector value $kl$ (b) with an asymmetric ($m=2$) configuration for $k_{so}=0.1\pi$, (c) with a symmetric ($m=3$) configuration for $k_{so}=0.15\pi$. }
\end{figure}

In the absence of a magnetic field ($B=0$), RSOC effects may be clearly seen in Fig. 3(a), which illustrates variations of the conductance $G$ with $k_{so}=0$ (solid line) and for $k_{so}=0.3\pi$ (dash line). From Fig. 3(a), we can see that the phase of the electrons in the absence of the RSOC is different from the case where the RSOC is present. In this case, the phase difference is half a period showing that the RSOC can also lead to a large conductance oscillation in our model. It can be seen that this phase also plays an important role in the poles and zeros of the conductance. We show the conductance of a polygonal ring with asymmetric leads for describing the effect of the RSOC in Fig. 3(b), which illustrates the variations of the spin-dependence conductances $G_{\uparrow}$ (solid line) and $G_{\downarrow}$ (dash line) with non-zero RSOC. There are two aspects of the RSOC influence on the conductances that deserve mention. Firstly Fig. 3(b) illustrates that the RSOC results in spin polarization. The reason is that the RSOC effects, although controlled by a perpendicular gate voltage, can be seen as being due to an in-plane effective magnetic field, which is momentum dependent. The interference between clockwise and counterclockwise transmission of the spin electronic wave functions can be controlled by the AC phase induced by the RSOC. Therefore, the AC phase depends on the route of the transmitted electron in the polygon, which can modify the intrinsic interference caused by the geometrical structure and results in spin polarization and AC oscillation in the conductance.
\

Secondly, we note from our calculation that the spin-dependent conductances have the same values when the output leads are in the axially symmetry positions. For example, $G_{m=2}=G_{m=4}$ and $G_{m=1}=G_{m=5}$ in a regular hexagon. The reason is that the RSOC in a polygonal ring does not break time reversal symmetry with the result that the spin up and spin down conductances have the same value when the output lead is located at symmetric positions with respect to the input lead (see Fig. 1). From Fig. 3(c), we can see that the degree of spin polarization of the transmitted electrons can be controlled by the strength of the RSOC. Since the RSOC does not break time reversal symmetry we have $G_{0\rightarrow m}=G_{m\rightarrow0}$. However, the magnetic field in our model breaks time reversal symmetry. Further transport properties controlled by the AB effect and the AC effect in our model are discussed below where both RSOC and the magnetic field are considered.

\begin{figure}
\centering
\includegraphics[scale=0.6]{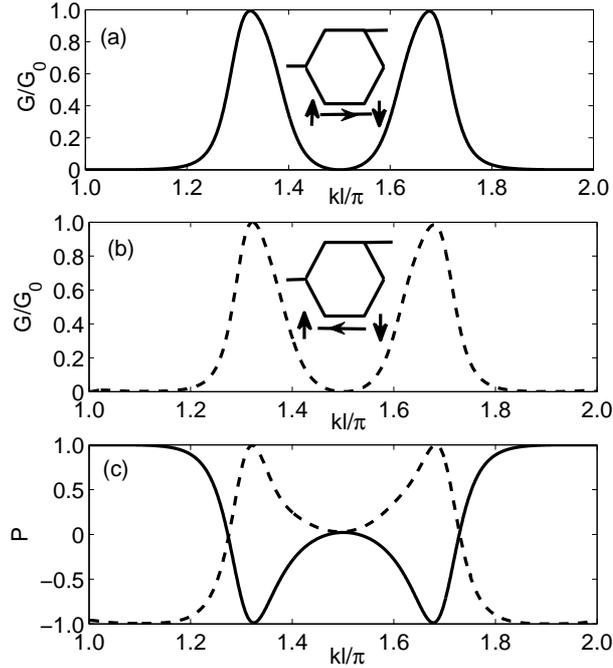}
\caption{\label{fig_1}(a) Conductance $G_{\downarrow}$ versus the wave-vector $kl$ for $G_{0\rightarrow m}$ when the input electrons have $s=\uparrow$ and $N=6$, $m=2$. (b) Conductance $G_{\uparrow}$ versus the wave-vector $kl$ for $G_{m\rightarrow0}$ when the input electrons have $s=\downarrow$ and $N=6$, $m=4$. (c) the efficiency of a spin splitter versus the wave-vector $kl$ for cases (a) and (b). Other parameters are $B=0.019~T$ and $k_{so}l=0.45\pi$.}
\end{figure}
\
The spin-dependent conductances in the asymmetric $N=6$, $m=2$ system for non-zero RSOC are shown in Fig. 4(a) where we use the parameters of an InAlAs/InGaAs heterostructure as a specific example.\cite{18} The Fermi wavelength is $\lambda_{F}\approx31.4$~nm and $k_{0}\approx0.2$~nm$^{-1}$, which is same as used in Ref. 21, and the incident electrons are taken to be polarized. The conductances with non-zero RSOC vary periodically as shown in Fig. 4(a). In Fig. 4(a), when the input electrons have $s=\uparrow$, we can see that the spin direction of the transmitted electron is down when $kl\simeq1.32\pi$, $N=6$, $m=2$. When the input electrons have $s=\downarrow$, as in Fig. 4(b), however, we can see that the spin direction of the transmitted electron is up when $kl\simeq1.32\pi$, $N=6$, $m=4$. From Fig. 4(a) and Fig. 4(b), we can see that the shapes of the curves are identical.
\

In order to describe more clearly the spin-dependent transport properties in a polygonal ring, the polarization, $P=\frac{G_{\uparrow}-G_{\downarrow}}{G_{\uparrow}+G_{\downarrow}}$, was calculated. The results are plotted in Fig. 4(c) for the case where the RSOC is non-zero and the output lead is located at an asymmetric position. It is interesting to note that when the input electrons are polarized, the device can be used for spin reversal from spin up to spin down for the $N=6$, $m=2$ case, and from spin down to spin up for the $N=6$, $m=4$ case. We can also note that the $N=6$, $m=4$ case is the same as the case where we exchange the source and the drain for the $N=6$, $m=2$ case. That is, for a fixed magnetic field and RSOC, when the electron flows from the left lead to the right lead, our device can produce a spin flip from spin up to spin down[See the inset figure in Fig. 4 (a)]. However, when the electrons flow from the right lead to the left lead, our device can produce spin flip from spin down to spin up[See the inset figure in Fig. 4 (b)]. That means that we can conveniently reverse the spin polarization of the transmitted electrons without changing the strength of the RSOC and the magnetic field.
\

\begin{figure}
\centering
\includegraphics[scale=0.52]{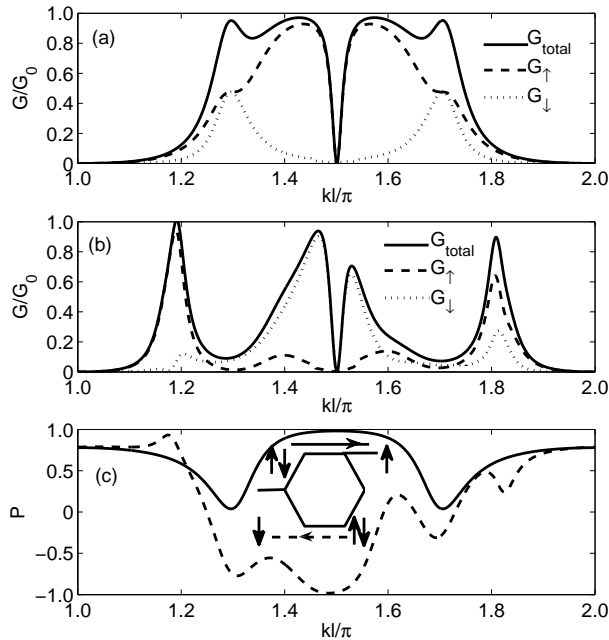}
\caption{\label{fig_1} Input electrons are unpolarized. (a) Conductance $G$ versus the wave-vector $kl$ for $N=6$, $m=2$. (b)Conductance $G$ versus the wave-vector $kl$ for $N=6$, $m=4$. (c) The efficiency of the spin splitter for $N=6$, $m=2$ (solid curve) and $N=6$, $m=4$ (dash curve) as a function of the wave-vector values. Other parameters are $B=0.027~T$ and $k_{so}l=0.125\pi$.}
\end{figure}

The variation of the conductances in a hexagonal ring as a function of the wave-vector are shown in Fig. 5(a) and Fig. 5(b) for fixed $B=0.027~T$ and $k_{so}l=0.125\pi$. The efficiency of the spin splitter for an asymmetric system with finite magnetic field and RSOC is shown in Fig. 5(c). Input electrons are taken to be unpolarized. We can see that the spin direction of the transmitted electrons is up when $kl\simeq1.46\pi$, $N=6$, $m=2$. However, the spin direction of the transmitted electron is down when $kl\simeq1.46\pi$, $N=6$, $m=4$ [See the inset figure in Fig. 5 (c)]. Therefore, the values of the RSOC and the magnetic field in the polygonal ring can be used as control parameters for a spin filter. In our model, the phase of the transmitted electrons depends on four factors. The first is the initial phase of the electron injected into the polygon. The second is the phase difference between upper and the lower arms which can be controlled by the location of output lead. The third and the fourth are the AB phase and AC phase which are induced by the magnetic field and RSOC effect, respectively. The last two factors can also be used to control the spin polarization. In our calculations, the phase is different for spin up and spin down electrons, which results in different values for the spin-dependent conductances. Therefore, our model polygon can be designed as a spin filter device. Because RSOC satisfies time reversal invariance the same spin transport properties are obtained when the output lead is located at either of a pair of axially symmetric positions. However, the magnetic field term violates time-reversal invariance. Consequently, when the left lead is connected to the source, our model can be used as a spin filter for spin up, for a specific set of parameters, and for the same parameters when the left lead is connected to the drain, the hexagonal ring can be used as a spin filter for spin down.
\

These results are expected to be useful for device design and quantum computation in the future.
\section{Summary}
In this paper, using a standard 1D quantum network method, quantum spin transport is investigated in a polygonal quantum ring with two leads. The effects of both a magnetic field and RSOC are taken ito account. The spin-dependent conductances can be calculated analytically using Griffith's boundary conditions. Transmission probabilities and polarizations are determined by both the magnetic field and the RSOC, and it is demonstrated that both the magnetic field and the RSOC can be used to control the phase of the transmitted electrons. In addition, the magnetic field in our model can break time reversal symmetry, with the result that the spin states of electrons transmitted from the source to the drain are different from the case where we exchange the source and the drain. We have also studied the conductance as a function of the wave-vector values when both the magnetic field and RSOC are considered, and have found that such a device can be a useful and effective means for controlling the spin polarized current.

\section*{Acknowledgement}

This work was supported by the National Natural Science Foundation of China under Grant No. 61176089, the Natural Science Foundation of Hebei Province under Grant No. A2011205092, and the Foundation of Shijiazhuang University under Grant No. XJPT002. We are also very grateful to Professor N. E. Davison for the enhancement of the English writing.


\end{document}